\documentclass[a4paper,10pt]{article}
\usepackage[utf8x]{inputenc}
\usepackage{multirow}
\usepackage{hyperref}
\title{A new Koide tuple: strange-charm-bottom.}
\author{A. Rivero  \thanks{\texttt{al.rivero@gmail.com}}}

\begin{document}

\maketitle

\begin{abstract}
With the negative sign for $\sqrt m_s$, the quarks strange, charm and
botton make a Koide tuple. It continues the previously found c-b-t
tuple and, more peculiar, it is cuasi-orthogonal to the original lepton
tuple. 
\end{abstract}

\section {A history of Koide sum rule}

In the late seventies, the empirical observation of a relationship between Cabibbo angle
and $\sqrt {m_d/m_s}$ drove an industry of models and textures for the quark mass 
matrix, simultaneusly to the advent of the third generation. Interesting
actors here are Wilczek and Zee, Fritzsch, and particularly Harari et al. \cite{Harari:1978yi}, who
 goes as far as to propose
a model that also implies a direct prediction of the $u,d,s$ masses: 
\begin{equation}\label{uds}
 m_u=0, \, {m_d \over m_s}={2-\sqrt 3 \over 2+\sqrt 3  }
\end{equation}
and, note, a Cabibbo angle of 15 degrees.

Eventually the model industry made use of preon models, a technique popularised
in Japan by Terazawa, who also proceed to suggest some more complex sum rules between
quotients of square roots, some of them coming from GUT or preon models, some of them 
being purely empirical. And both GUT based and preon based models allowed
eventually to extend the equations to
the lepton sector, which was more promising, given the high precision of the 
mass of electron and muon.

So, in the early eighties Koide suggested some models \cite{koide, koide2, Koide:1983qe} able
to predict a Cabibbo angle exclusively from the lepton sector, and then he found
that some of the equations also predicted a relationship between the three charged
leptons: 

\begin{equation}\label{koideq}
 { (\sqrt m_e + \sqrt m_\mu + \sqrt m_\tau )^2\over m_e + m_\mu + m_\tau } = \frac 32
\end{equation}

Really Koide's equation predicted the mass of the tau lepton before the 
correct measurement of its current value,
and still now it is exact inside one-sigma levels. But at the time of the proposal,
the experimental measurement of tau was not so good and it was mistaken in some
percentage. Thus the ``sum rule'' was kept sleep until a re-evaluation of the tau mass
vindicated it. At that time, in the mid nineties, the field has evolved and the 
original goals were not in modern check lists, so the formula got a new start.  

Foot \cite{Foot:1994yn} suggested to read the relationship more geometrically, as if asking 
the triple of square roots to keep an angle of 45 degrees with the triple $(1,1,1)$. 

Esposito and Santorelli \cite{Esposito:1995bw} do an analysis of the renormalisation
running, and a first approachment to the question of fitting the neutrinos, if massive,
into a similar formula.

In 2005, an informal online working group\footnote{Around www.physicsforums.com and
some blogs, after some headstart in s.p.r newsgroup. Besides the authors of the 
referenced papers, other strong online contributors as Hans de Vries or Dave Look 
provided alternate insights and even programmatic analysis tools} 
took the task of evaluating low energy
mass formulae and its current validity, and of course Koide formula emerged again
here. An incomplete review of the formula was done in \cite{Rivero:2005vj}, addressing the case
of zero mass, where the relationships (\ref{uds}) are recovered. The neutrino case
was reevaluated with most modern data \cite{Brannen:2010zza,Brannen:2010zz}. Eventually the 
reference to Koide sum rule for neutrinos did its way into the standard
literature on PMNS parameters. 

A byproduct of the online effort was to rewrite again the equation following Foot's idea,
allowing a phase angle to parametrise the rotation around $(1,1,1)$. 
\begin{equation}\label{fase}
m_k = M (1 + \sqrt 2 \cos (2 k \pi /3 + \delta_0))^2
\end{equation}
It is usual to absorb the permutation ambiguity of Foot's cone in this phase $\delta$, by the combination
of change of sign, plus rotations of 120 degrees. 

It is intriguing that for charged leptons $M\approx 313$ MeV, typical of constituent 
quarks or of QCD diquark strings. But more important is that this parametrisation
clarifies the use of negative signs in the square roots of the masses. This was
important to build the neutrino tuples, and it is relevant for the new tuple that
we are presenting in this paper.

Inspecting (\ref{fase}), you can see that there are two ways to produce a degenerated
pair: with $\delta=0$ or with $\delta=\pm\pi/12$, and you can use them to produce one or
other hierarchy of neutrinos. Being the phase of the charged leptons $\delta_l$, C Brannen
proposed \cite{Brannen:2010zza}  a phase $\pi/12+\delta_l$ to match known bounds, and  M. D. Sheppeard
proposed \cite{md} $-\pi/12+\delta_l$ as a match to the results of MINOS experiment.

The PhD thesis of François Goffinet \cite{Goffinet08}, in 2008, revisits most of the 
old concepts, and then including neutrinos and the possibility of negative roots, as well
as the idea of generalising the formula to all the six quarks in a single sum. This last
possibility has been also reviewed by \cite{Kartavtsev:2011jt}.

\section {Current advances}

Very recently Rodejohann and Zhang \cite{Rodejohann:2011jj} recognised the 
possibility of fitting Koide formula to quark triplets not of the same charge,
but of nearby mass: they suggested fits for the low mass quarks $uds$ with
current values of the mass quotients, and more importantly for this note,
they suggested a very good fit for the heavy quarks $cbt$. This can be
readily verifyed from current data from \cite{pdg}. With $m_t=172.9 \pm 0.6 \pm 0.9$,
$m_b=4.19_{−0.06}^{+0.18}$, and $1.29_{−0.11}^{+0.05}$ GeV, the central values
give in \ref{koideq} a LHS value of about $1.495$, very close to the
required $1.5$. Further analisis of it, including renormalisation group,
can be seen in \cite{Kartavtsev:2011jt}.

It is interesting to try to produce all the masses
from the two upper ones. We can solve (\ref{koideq}) as
\begin{equation}\label{descent}
m_3(m_1,m_2)= \left((\sqrt m_1 + \sqrt m_2 ) \left(2- \sqrt{3+6 {\sqrt{m_1 m_2} \over 
(\sqrt m_1+\sqrt m_2)^2}} \right)\right)^2
\end{equation}
and use it to iterate. We get the descent:

\begin{center}
$
\begin{array}{lcl}
m_t &=& 172.9 \mbox{ GeV} \\
m_b &=& 4.19 \mbox{ GeV} \\
m_c(172.9,4.19) &=& 1.356 \mbox{ GeV} \\
m_s(4.19,1.356) &=& 92 \mbox{ MeV} \\
m_u(1.356,0.092) &=& 0.036 \mbox{ MeV} \\
m_d(0.092,0.000036) &=& 5.3 \mbox{ MeV}  
\end{array}
$
\end{center}

The main point in this descent is that we have produced a tuple not yet in the literature,
the one of strange,charm,and bottom. How is it?

Closer examination shows that the reason of the miss is that in order to meet (\ref{koideq}), the value of
$\sqrt{s}$ must be taken negative. But this is a valid situation, according Foot interpretation and the
parametrisation (\ref{fase})

Of course, once we are considering negative roots, the equation (\ref{descent}) is not the only possible matching. But
the possibilities are nevertheless reduced by the need of a positive discriminant in the equation and by
avoiding to come back to higher values, above the mass of the bottom quark. Also, one we have recognised the sign
of $\sqrt{s}$, the validity of the two next steps in the descent, up and down, is unclear. We will come back to
these two quarks in the next section.

Another important observation is that  $(-\sqrt m_s,\sqrt m_c,\sqrt m_b)$ is on the opposite extreme of Foot's cone
respect to $(\sqrt m_\tau,\sqrt m_\mu, \sqrt m_e)$, making an angle of almost ninety degrees.

Furthermore, the parameters of mass and phase of this quark triple\footnote{For other quark triples we
have not found any obvious hint; for the charm,bottom,top, we get $M=29.74 GeV, \delta=0.0659$} seem
to be three times the ones
of the charged leptons: we have $M_q=939.65$ MeV and $\delta_q=0.666$, while in the 
leptons $M_l=313.8$ MeV and $\delta_l=0.222$, about $12.7$ degrees.

We could take seriously both facts and use them to proceed in the reverse way: 
take as only inputs the mass of electron and muon, then recover $M_l$ and $\delta_l$, multiply times three
to get the parameters of the opposite tuple and then the masses of strange, 
charm and bottom, and then use the ladder up and down to recover the previous table. It is impressive: 

\begin{center}
\begin{tabular}{l|l|l}
Inputs & Outputs\\
\hline
$m_e= 0.510998910 \pm 0.000000013$ & $m_\tau=1776.96894(7)$ MeV \\
$m_\mu=105.6583668 \pm 0.0000038 $ & $m_s=92.274758(3)$ MeV\\
$M_q=3M_l$ & $m_c=1359.56428(5)$ MeV\\
$\delta_q=3\delta_l$ & $m_b=4197.57589(15)$ MeV\\
 & $m_t=173.263947(6)$ GeV\\
\cline{2-2}
 & $m_u=0.0356$ MeV \\
 & $m_d=5.32$ MeV
\end{tabular}
\end{center}

Still, the results are a bit higher respect to the descent from the top quark. Actually, we don't
have an argument to keep the factor three between leptons and quarks, except that the physical
situation seems to be a small perturbation respect to a case where it is a bit more justified,
the full orthogonal case with $\delta_l=15^{\circ} $, $\delta_q=45 ^{\circ}$. Lets look at it now.

\section {The $m_e=0$, or $m_u$=0, limit}

Our purpose is to think of the empirical triples as a rotation from a more symmetrical situation, 
keeping the sum of the masses (which
is the modulus square of Foot's vector) constant. Our candidate
unperturbed state is the one who has the mass of the electron equal to zero.

It is precisely when $\delta=15^{\circ}$ in the parametrisation (\ref{fase}), that one of the
masses becomes cero and the mass tuple is
\begin{equation}
 m_{15}= \left( 3 (1+ {\sqrt 3 \over 2}) M, 0 , 3 (1 - {\sqrt 3 \over 2}) M\right)
\end{equation}
You can notice that this is equal to the result (\ref{uds}) above. 

The state in the opposite generatrix of the cone is given by a phase of $-165$ degrees, which is,
modulus the $120^{\circ}$ and sign ambiguities, $45$ degrees, ie three times the other phase. Our argument
is thus that for a small perturbation this factor can be kept at least at first order, if not better.

Now, lets contemplate the components of this orthogonal state:
\begin{equation}
 m_{45}=\left(  (1 - {\sqrt 3 \over 2}) M', 4M', (1+ {\sqrt 3 \over 2}) M'\right)
\end{equation}

There are obviously orthogonal (remember that we take $-sqrt(s)$) for any values of M and N', but
for $M'=3M$ there is an extra symmetry, or an extra degeneration of levels if you wish. So again
we incorporate this relationship, and its approximate validity under perturbation, as a postulate.
If you look at the empirical data above, you will see that $M_q/M_l \approx 2.99$.

So we identify $m_{45}$ as the quark triplet of strange, bottom and charm. Now we can proceed in the
same way that before, to produce from bottom and charm the mass $m_t$, and from strange and charm
the mass $m_u$, and next $m_d$. Just to put numbers in, and see how far we are from the perturbed
state, lets fix the only parameter $M$ to $313.86$ MeV. Then we get $m_s=m_\mu=126.1$ MeV,
$m_c=m_\tau=1757$ MeV, $m_b=3766$ MeV and $m_t=180$ GeV.

For $m_u$ we need to use $\pm \sqrt m_s$, and then we have two possibilities in (\ref{descent}). But the
alternative with the minus sign produces a zero discriminant, and then
the same mass than $m_b$. We can either to claim a halt here, and disregard the possibility to 
estimate $u$ and $d$, or to choose the plus sign.

In this case, the triple $u,s,c$ has the same solution that the lepton triple, and then $m_u=0$.

And the final ladder in the descent, $d,u,s$ with $m_u=0$ again does not allow for negative roots,
just look at (\ref{fase}) to verify it. So is again the same proportion, and now really
it is the relationship (\ref{uds}), with $m_d= m_s {2-\sqrt 3 \over 2+\sqrt 3  }= 9.05$ MeV. 

The reproduction of equation (\ref{uds}) in this formalism justifies the descent with positive sign;
but it is not fundamental for the rest of the work.

A last remark: Which is the mass of the pion in this limit? With $m_d$ still not null and higher that the
perturbed $m_u+m_d$, it seems even a bit higher than usual. But it would be interesting to
find an unperturbed model 
where $m_\pi=m_\mu$. Then, given that $m_e=0$, the charged pion would be stable.

\section {Conclusions}

 
We have shown that by taking Koide equation, parametrised in Foot's
cone, and then allowing determinate negative square roots, a new triplet
of quarks can be legitimally added to the collection of fermion sum rules.

Furthermore, we have seen that this triplet, strange-charm-botton, is almost
orthogonal to the original triplet of charged leptons; we have argued that it is 
valid to accept the proporcionality constants from this orthogonality
and to translate the parameters from one triplet to another. This allows
to calculate the masses of the quark triplet and then, applying Koide 
formula, also the mass of other quarks, and 
particularly \cite{Rodejohann:2011jj} the
mass of the top quark.

We have not done the effort of exploring the variation of the matchings
along renormalisation group runnings, as the general insights from other
works on Koide equation will apply here too. But we raise the suspicion
that the matchings are to be interpreted in the infrared or at last
at low energy, because the conspicuous role of mass scales usually
associated to QCD and chiral symmetry breaking: note the above empirical
values for the basic mass of lepton and quark triplets, $313.8$ and 
$939.7$ MeV.

\end{document}